# Explorable Ideas: Externalizing Ideas as Explorable Environments

Euijun Jung
Human-centered Computer Systems Lab, Seoul National University
Republic of Korea
euijun.jung@hcs.snu.ac.kr

Jingyu Lee
Human-centered Computer Systems Lab, Seoul National University
Republic of Korea
jingyu.lee@hcs.snu.ac.kr

Minji Kim
Human-centered Computer Systems Lab, Seoul National University
Republic of Korea
minji.kim@hcs.snu.ac.kr

Youngki Lee
Human-centered Computer Systems Lab, Seoul National University
Republic of Korea
youngkilee@snu.ac.kr

## ABSTRACT

Working with abstract information often relies on static, symbolic representations that constrain exploration. We introduce Explorable Ideas, a framework that externalizes abstract concepts into explorable environments where physical navigation coordinates conceptual exploration. To investigate its practical value, we designed Idea Islands, a VR probe for ideation tasks, and conducted two controlled studies with 19 participants. Results show that overview perspectives foster strategic breadth while immersion sustains engagement through embodied presence, and that seamless transitions enable flexible workflows combining both modes. These findings validate the framework's design considerations and yield design implications for building future systems that treat information as explorable territory across creative, educational, and knowledge-intensive domains.

**ACM Reference Format:**
Euijun Jung, Jingyu Lee, Minji Kim, and Youngki Lee. 2025. Explorable Ideas: Externalizing Ideas as Explorable Environments. In . ACM, New York, NY, USA, 17 pages. https://doi.org/10.1145/nnnnnnn.nnnnnnn

## 1 INTRODUCTION

Working with abstract information—whether for creative ideation, knowledge exploration, or reflective sensemaking—has long relied on symbolic representations such as text, diagrams, or maps [11, 77, 89]. These forms externalize ideas for observation, requiring users to read and interpret symbols to make sense of evolving structures [7, 17]. While effective for capture and organization, such observational tools remain visually static and cognitively detached from how thinking and discovery naturally unfold.

Recent findings in cognitive psychology and HCI have demonstrated that walking and embodied interaction can stimulate creative thinking by promoting cognitive flexibility and reducing mental fixation [35, 62, 91]. Embodied movement supports associative thinking, even short walks significantly increasing the number of ideas and originality compared to seated conditions [3, 62]. However, most systems remain sedentary in their interaction design, failing to leverage the cognitive benefits of movement.

This work proposes *Explorable Ideas*, a concept of representing abstract information as explorable environments, built on the foundational work in information landscapes [54] and inhabited information spaces [13, 80]. The core aspiration is achieving effective *spatial-semantic coordination*, where users develop meaningful relationships between physical navigation and conceptual exploration through embodied interaction [19, 32]. Rather than treating ideation contexts as static information displays or using virtual environments merely for atmospheric effect [24], explorable ideas enable users to inhabit their evolving creative landscape. The environment dynamically evolves as new ideas emerge, creating responsive landscapes that adapt to users' creative development while supporting both strategic overview and immersive exploration through different spatial perspectives [15, 98].

As the first realization of the *Explorable Ideas* framework, we designed *Idea Islands* as a probe for ideation tasks in VR. In this system, each conceptual theme is represented as an island, and each individual idea appears as a tree on that island. This spatial metaphor makes ideas navigable: users can step back to see the overall balance between themes, move inside a category to focus on its details, and sense gaps or connections between ideas. As new ideas are generated, the islands grow in real time through LLM-assisted semantic organization, allowing the landscape to evolve naturally with the flow of ideation. The purpose of this probe is to show how abstract information can be externalized in an explorable form and to examine what design choices best support information exploration.

Through two controlled studies using the Idea Islands probe with 19 participants, we establish the foundations of the Explorable Ideas framework for spatial information exploration systems. Our investigation addresses two fundamental questions: how users engage differently when treating explorable ideas as spectator versus inhabited experiences, and how they navigate transitions between and within semantic levels to sustain exploration. Study 1 demonstrates that spectator perspectives foster reflective assessment and planning across conceptual landscapes, while inhabited experiences enable persistent engagement within focused domains through embodied presence. Study 2 reveals how users naturally develop integrated workflows, leveraging seamless transitions between these modes for enhanced creative agency. Our findings establish three core design implications: the need for aesthetic and carefully sized props that sustain exploration, the critical balance between embodied engagement and cognitive efficiency in transition design, and the importance of *spatial-semantic correspondence* as a promising evaluation criterion.

Table 1: Representative use cases of Explorable Ideas and their information structures.

| Use Case | Information Structure | Spatial Encoding | Metaphor Example |
| --- | --- | --- | --- |
| **Ideation (Section 3)** | Themes → ideas | Categorical containment | Islands, trees, props |
| **Lifeline Archiving** | Timeline → events | Sequential pathway; chronological continuity | Timeline path; galleries |
| **Retrospective Photobook** | Chapters → media items | Ordered/ thematic grouping; album-like browsing | Bookshelves; pages/photos |
| **Knowledge Exploration (Figure 1)** | Domain → topic → concept → references | Containment; connecting lines for cross-references | Library halls; shelves; books |

Our contributions are:
- We introduce the *Explorable Ideas* framework as a design concept for externalizing abstract information into explorable environments.
- We present *Idea Islands*, a probe system that realizes this concept for ideation tasks, externalizing ideas as embodied and explorable forms through dual-scale interaction and natural metaphors.
- We derive the design implications for explorable ideas systems through empirical studies with *Idea Islands*.

## 2 EXPLORABLE IDEAS

*Explorable Ideas* is a concept for organizing information into the environment of intuitive structure. An individual piece of information is reified into the form of environments or props, allowing users to directly *experience* rather than merely observe information [7, 17, 54]. By mapping semantic relations into spatial layouts, the framework enables users to grasp conceptual structures through embodied navigation, where physical movement corresponds to conceptual exploration [19].

### 2.1 Use Cases

This concept can be applied to various practices, such as ideation, archiving, retrospection, and knowledge exploration, extending principles from spatial information design [13, 80] to experiential contexts [52, 54, 82]. Table 1 outlines representative use cases and their corresponding information structures.

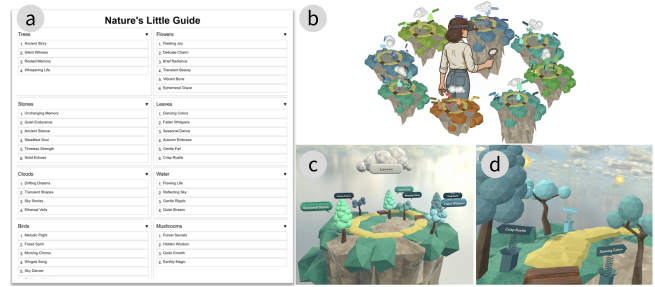

Figure 1: Illustration of *Knowledge Islands*: (a) Structured data created through a web-based interface; (b) conversion of the data into island-based spatial layouts; (c) an overview of the resulting landscape; (d) an inside view along a navigable path.

Figure 1 illustrates the case of *Knowledge Islands*. Here, a personal knowledge base (e.g., notes managed in tools like Obsidian or Notion [60, 61]) is externalized into a spatial landscape: each topic is represented as an island, individual notes appear as trees or props, and cross-references can be encoded as visible connections across the terrain. This enables the coordination between semantic and spatial territories, where users walk through their knowledge base, perceiving clusters of related concepts and discovering gaps through spatial distance, leveraging spatial reasoning for information exploration [40, 88]. Similarly, other practices can be instantiated in explorable form—for instance, a lifeline archive spatialized as a chronological pathway supporting temporal navigation, while a photobook retrospective is experienced as albums and pages that accommodate both overview browsing and conversion into live environments of corresponding events.

### 2.2 Requirements

Realizing *Explorable Ideas* requires designers to satisfy three intertwined requirements:

- **Embodiment** — ideas should be reified as places that can be walked through and inhabited, rather than passively observed, leveraging embodied cognition principles [94].
- **Exploration** — the system must allow users to flexibly shift between global overview and local immersion, supporting both breadth-first exploration and depth-focused elaboration through multi-scale interfaces [15, 98].

- **Semantic Fidelity** — spatial metaphors should not come at the cost of informational completeness; representations must preserve and convey the evolving meaning and detail of ideas [7].

## 2.3 Design Considerations

Realizing *Explorable Ideas* in practice involves several design considerations. We organize them as guiding questions that connect abstract concepts to concrete design choices.

**How to structure information?** A foundation of building *Explorable Ideas* is determining the depth and form of hierarchy. Information exploration rarely produces isolated elements; users expect to move between broader themes and finer details in a structured progression [15, 73].

An individual piece of information (which we call an IDEANODE) is always situated within a hierarchy. The node has dual roles, both as an object in a structure and an environment to inhabit. From an external view, IDEANODE appears as a discrete node within a larger structure, but when the user enters it, the same node is transformed into an explorable environment. Its internal subnodes are then instantiated as props or landmarks inside this environment, preserving the containment relation while shifting the user's perspective from abstract hierarchy to embodied space.

The key design question is how finely to subdivide: excessive granularity can fragment meaning and overburden navigation, while overly coarse grouping risks obscuring important semantic distinctions [20]. Equally important is the mechanism of structuring—whether categories are generated automatically, manually curated, or shaped through hybrid editing workflows [11, 25, 30, 77]. The degree of user control here directly affects both cognitive load and sense of ownership in the evolving space [67, 76].

**How to represent information?** Once structured, ideas must be reified as perceivable elements in the environment with proper appearance, location, and scale.

Visual appearance can take the form of symbols, textual panels, or hybrid metaphors, each trading off between immersion and informational detail [7]. Spatial location encodes semantic relations: related categories can be placed in close proximity, while temporal or narrative order can be mapped along a linear axis, supporting situated exploration [22]. Scale further shapes the experience—lifesized elements enhance presence but may cause clutter, whereas miniature forms provide an overview but risk detachment [86, 98]. Maintaining walkability and legibility is essential so that physical navigation remains aligned with conceptual relations [1, 19].

The representational choices should adapt to how an IDEANODE is experienced. When presented as part of an external environment, emphasis should be placed on the relative positioning of hierarchy. When inhabited internally, however, the same IDEANODE must function as an environment where local density, distribution, and skewness of sub-elements shape how information is perceived and navigated. For example, a sparsely presented idea may highlight conceptual gaps, whereas a densely clustered one motivates users to reorganize [17].

**How to support transition within and between IDEANODES?** Experiences of *Explorable Ideas* are realized not only in static representation but in the process of moving through IDEANODES.

Within an environment, visibility policies—always-on panels versus perspective-dependent reveal—determine whether users passively scan or actively seek content. Between environments, transitions critically influence continuity: instant teleportation affords efficiency but can break immersion, whereas embodied traversal reinforces a sense of place and narrative progression [14, 51]. To preserve coherence, global overviews and local immersions must remain consistent, retaining awareness of the overall structure [15].

## 2.4 Related Work

We examine four foundational research areas that inform the Explorable Ideas framework: spatial representation of information, embodied cognition principles, scale and perspective effects in virtual reality environments, and responsive content organization.

*2.4.1 Spatial Representation of Information.* The foundational vision of transforming abstract information into navigable spatial environments emerged from Muriel Cooper's pioneering concept of Information Landscapes [54]. This paradigm proposed that information could be given spatial form and dimensional meaning, with early systems demonstrating how perspective and spatial positioning could foster new ways of understanding abstract information [52, 79, 82]. The concept of Inhabited Information Spaces [80] further advanced this vision by establishing design principles for creating navigable information environments [22] and developing spatial interaction models for large virtual worlds [4]. The visual design of these information landscapes significantly influences cognitive processing, as spatially externalizing thoughts alters mental operations by offloading working memory constraints [40] and grounding abstract concepts in tangible experiences [17, 42]. Metaphorical structures represent complex conceptual relationships effectively when aligned with users' spatial-conceptual schemas [7, 48, 70], influencing the reactiveness to feedback and collaborative dynamics [83–85]. VR environments provide a potent medium for realizing spatial information contexts through contextual priming and immersive environmental design [24, 50]. Research demonstrates that immersive spaces can trigger cognitive associations and support creative performance [5, 57], with recent systems featuring AI-powered spatial representations [95], mind-mapping platforms [78], and environmental manipulation techniques [6, 38]. However, existing VR approaches often treat environments primarily as *atmospheric contexts* rather than *semantic territories*, limiting their potential for content-driven spatial exploration. Our Explorable Ideas framework addresses this by creating environments where physical navigation directly corresponds to semantic discovery, integrating spatial legibility with content exploration.

*2.4.2 Embodied Cognition and Virtual Reality.* Embodied information processing emerges from brain-body-environment interactions, with physical action fundamentally shaping thought and meaning [19, 32, 88]. Empirical evidence supports this through sensorimotor processes in higher-order thinking [53, 94], with walking increasing divergent thinking by up to 81% [62] and hand gestures enhancing creative performance [35]. VR environments offer unprecedented opportunities for embodied interaction, enabling spatial reasoning and metaphorical enactment that transcend physical world constraints, with unique affordances enabling new forms of spatially-aware interaction [37, 71]. Wang et al. [91] show that physically enacting metaphors in VR can measurably enhance cognitive output. The impact of embodied interaction extends to therapeutic applications where mindfulness-based spatial interaction promotes well-being [27, 72], and design research where VR environments fundamentally alter cognitive processes through embodied spatial reasoning [10, 12, 44, 97]. Our work extends these insights by investigating how body-scale interaction within structured information environments can support deliberate conceptual discovery through embodied navigation within semantic landscapes.

*2.4.3 Scale and Perspective Effects in Virtual Reality.* The relationship between scale and cognitive processing has profound implications for information exploration. Different scales afford distinct cognitive modes for information exploration, though scale transitions impose working memory burdens requiring careful management [15, 63]. Discrete scale transitions provide clearer cognitive boundaries than continuous zooming [63], with malleable interfaces enabling adaptive workflows [56]. VR environments introduce unique opportunities and challenges for multi-scale interaction that alter spatial cognition and information processing [23], with the World-in-Miniature (WIM) paradigm establishing that miniature representations enable strategic overview while maintaining spatial coherence with full-scale environments [81, 87, 90]. Research has produced sophisticated strategies for managing scale transitions, including multi-scale teleportation techniques supporting different cognitive engagement modes [92], walking-oriented navigation with virtual body resizing [41], and systematic viewpoint transition techniques [45], though significant perceptual challenges arise when users are resized within virtual environments, as scale changes can disrupt spatial understanding and presence [65]. Our work builds on these insights by investigating how discrete scale transitions between overview and immersive exploration can support different phases of information discovery within seamless exploration experiences.

*2.4.4 Responsive Content Organization.* Information management systems have evolved from basic capture and categorization [26, 59] toward sophisticated real-time analysis and adaptive organization, establishing core paradigms through crowd-powered semantic modeling platforms like IdeaHound [77], real-time content visualization systems such as TalkTraces [11], and collaborative coordination platforms [18, 39, 74, 89]. The advent of Large Language Models has marked a paradigm shift, enabling intelligent content generation and adaptive organization [16, 34, 36, 69, 96]. Research on human-AI collaboration has demonstrated increased productivity and content diversity through digital facilitation mechanisms [31, 76, 93], though studies caution that poorly implemented automation can constrain exploration and that preserving human agency remains crucial [67]. Our approach addresses this by integrating AI-powered semantic organization with spatial embodied exploration, creating immersive information landscapes benefiting from intelligent, real-time organization.

# 3 IDEA ISLANDS: A PROBE FOR EXPLORABLE IDEATION

*Idea Islands* is the first realization of the *Explorable Ideas*, designed as a tool for ideation in a VR environment. Ideation is inherently cognitively demanding [58]. It requires participants to both generate new ideas and sustain evolving connections among them [28]. Our probe aims to alleviate this burden by reifying abstract ideas into a spatial and experiential form, where information is not only transmitted but also embodied as a lived experience [42, 94]. In this setting, contents dynamically change as new ideas are produced, allowing users to navigate between global overviews and local elaborations.

We later study *Idea Islands* to uncover practical implications for how *Explorable Ideas* can be constructed in future systems. This probe serves both as a demonstration of the framework's principles and as an empirical ground to examine how spatialized, experiential representations can support creative thinking.

## 3.1 Idea Islands Design

*3.1.1 Spatial Artifacts and Visual Metaphors.* Figure 2 shows the overview of the island design. To address the challenge of representing information (Section 2.3), Idea Islands adopts metaphors of natural themes, which support creative ideations [21, 29, 47]. Each IdeaNode is represented as a perceivable prop that can be both a node in a hierarchy and an explorable environment when entered.

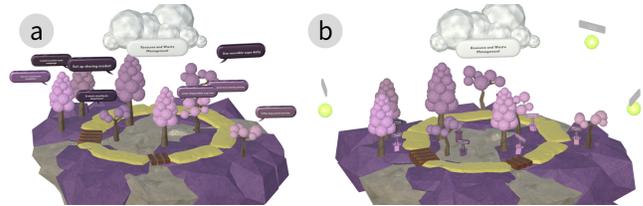

**Figure 2: Idea Island represents ideas and their category as trees and an island: (a) the mini-scale form; (b) the body-scale form of the integration mode (Study 2).**

Two artifacts encode the ideation hierarchy as IdeaNode:
- **Islands**: Represent conceptual categories as navigable domains.
- **Trees**: Represent individual ideas instantiated within domains.

The island–tree relationship provides a natural containment metaphor that is both intuitive and spatially legible.

Other artifacts serve as auxiliary elements to enhance context, readability, and continuity across scales:
- **Clouds**: Display category names above islands, providing semantic labels.
- **Pop-ups/Signposts**: Adapt idea content presentation to interaction scale (pop-ups for overview, signposts for immersive mode).
- **Orbs**: Provide simplified peripheral indicators of non-current categories, notifying users if ideas are categorized into the corresponding category [66]. Note that orbs are placed over the pathway or at the island edge by transition modes.

- **Pathways**: Establish walking guidance within islands and define where new ideas appear, supporting predictability and legibility [22].

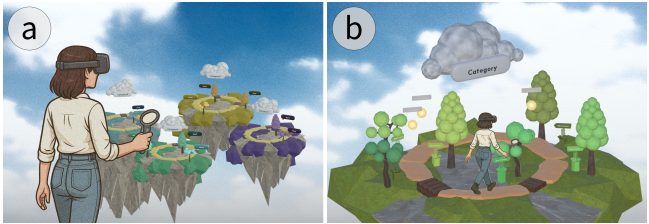

Figure 3: Dual-scale interface design: (a) mini-scale overview mode for rapid scanning and assessment; (b) body-scale exploration mode for immersive navigation within domains (Study 1).

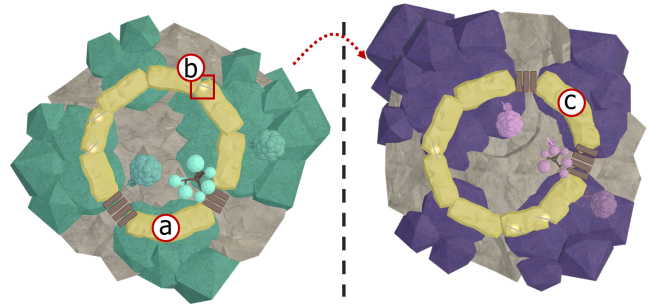

Figure 4: Walk-to-Teleport transition mechanism: (a) user's starting position within an island; (b) user walks toward and clicks near the target orb; (c) new island instantiated with pathway aligned to user's physical position for seamless entry.

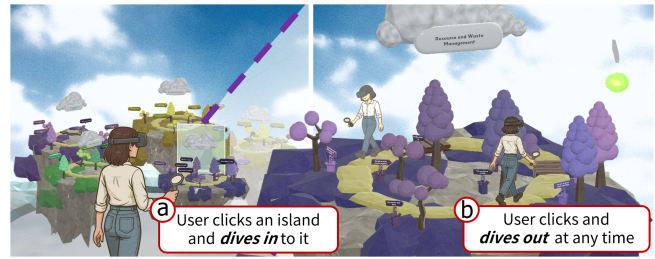

Figure 5: Dive-In/Dive-Out transition mechanism: (a) dive-in action from overview mode to body-scale exploration; (b) dive-out action returning to mini-scale overview mode with maintained spatial context.

*3.1.2 Dual-Scale Interface.* Idea Islands supports two complementary modes that reflect the dual role of IDEANODE. Each island is represented as a miniaturized overview for rapid scanning and a body-scale immersion for situated engagement (Figure 3). The *overview mode* positions users above a miniature landscape, enabling rapid contextual scanning, recognition of emerging categories, and efficient review of idea content through oriented pop-ups [15]. Here, all categories remain visible at once, supporting strategic breadth-first exploration. The *immersive mode* enlarges a chosen island to human scale, allowing users to walk among idea trees, read signposts, and dwell within a semantic space. We limited the visibility of each element depending on the user's perspective to sustain immersion. For example, idea content is revealed on signposts only when approached at a right angle. Pathways guide users through the environment, providing predictable routes and shaping encounter order. Other categories are reduced to peripheral orbs, minimizing distraction while preserving awareness of the broader context. To minimize physical collision that harms the spatial experience, we managed the positioning and size of the island, limiting the number of potential trees in a single category to eight.

*3.1.3 Transition Between Islands.* We developed two transition techniques that preserve spatial continuity while addressing the conflict between continuous virtual navigation and the restricted area available for real-world movement.

- **Walk-to-Teleport**: A user walks toward an orb representing the target domain (Figure 4.a-b); the current island fades out and the target island is instantiated surrounding them, rotated to align its pathway with the user's current position (Figure 4.c). This approach reinforces embodiment and preserves room-relative available spaces for motion.
- **Dive-In/Dive-Out**: In overview mode, users can *dive in* to enlarge a target island to body scale (Figure 5.a), or *dive out* with a single click to return to the overview (Figure 5.b). Room-relative remapping ensures effective use of tracking space across scales.

Together, these mechanisms balance efficiency and immersion: Walk-to-Teleport emphasizes embodied traversal, while Dive-In/Out provides rapid contextual switching.

*3.1.4 Dynamic Organization of Islands.* Idea Islands dynamically restructures the environment based on real-time speech input and LLM inference (see Algorithm 1 in Section A.3). Utterances are transcribed and categorized; if a suitable island exists, a new tree is instantiated, otherwise a new island is created. This adaptive logic allows the environment to grow in response to ideation, supporting both temporal sensitivity and semantic coherence. Animations and sound cues accompany instantiations to provide perceptible feedback, reinforcing user engagement.

## 3.2 Implementation Details

The system was implemented in Unity 2022.3.13f1, using WhisperX [2] for speech-to-text transcription and GPT-4o [33] for semantic organization and content summarization. The response time from user utterance to visual instantiation was optimized to be 3 seconds to maintain natural conversational flow during ideation sessions.

## 4 STUDY OVERVIEW

Building on the Explorable Ideas framework (Section 2.3), we conducted two controlled studies with 19 participants using our Idea Islands probe. These studies examine how users engage with explorable idea spaces both as spectators (overviewing structures) and as inhabitants (embodied exploration), and how they transition across and within semantic hierarchies. We focus on two research questions:

**RQ1**: How do users engage differently when treating ideas as a *spectator experience* versus an *inhabited experience*?

**RQ2**: How do users navigate ideas both within a single perspective and across transitions between global overviews and detailed explorations?

Study 1 (N=13) isolates dual-scale interface effects by comparing mini-scale overview and body-scale exploration modes, investigating how different spatial scales support temporal dynamics of ideation and what cognitive engagement patterns emerge. Study 2 (N=6) examines integrated systems where users construct workflows with seamless transitions, revealing natural exploration strategies and investigating how embodiment factors into transition design within spatial-semantic coordination environments. We collected mixed-method data, including creative performance metrics, experience ratings, spatial tracking, and qualitative interviews. Both studies received institutional review board approval.

## 5 STUDY 1: COMPARISON BETWEEN OVERVIEW AND IMMERSION MODES

This study examines how spatial scale shapes ideation. We compare *mini-scale overview* and *body-scale immersion* to reveal how different perspectives foster distinct cognitive modes and creative strategies.

### 5.1 Methodology

*5.1.1 Study Conditions.* We used a focused within-subjects design with two conditions to isolate the effects of spatial scale:

- **Mini-Scale Overview (MINI):** Provides an overview of the entire idea space.
- **Body-Scale Exploration (BODY):** Provides first-person immersive view on a single island.

*5.1.2 Participants and Procedure.* We recruited 13 participants (8 female, age M=23.85, SD=3.21) through university postings. Participants had varying ideation experience: 2 ideated 1-2 times daily, 6 ideated 1-2 times weekly, and 5 ideated less than monthly. Ten participants rated their ideation proficiency as 4+ on a 7-point scale. The VR experiences were as follows: 3 participants used VR twice, 5 used it three to four times, and 5 used it five or more times. Participants engaged in open-ended ideation on two topics: "How can we facilitate communication among people in campus lounges and cafes?" and "What can be done to encourage healthier habits among students?" Each participant completed both conditions in counterbalanced order. Sessions lasted 4-8 minutes, with participants notified at the 4-minute mark that they could continue or finish. At the start of the experiment, participants received task instructions and were introduced to using the artifact. For the BODY condition, participants were instructed to walk along paths (refer to *walk-to-teleport* for details in Section 3.1.3). When participants generated nine or more ideas within a single category with sufficient time remaining, they were guided to explore other categories. This guidance was provided to two participants (S1P8 and S1P10, both in body-scale conditions). Post-session measures included experience ratings on eight dimensions, including self-satisfaction and attractiveness (0-100 scales), spatial presence questionnaire [75], system usability scale [9], and dynamic evolution quality ratings. After two ideations, we conducted a semi-structured interview, focusing on how the different interfaces influenced their creative process and awareness of spatial artifacts. The experiment took place in a space larger than $33m^2$ where participants could move around safely. The experiment employed a VR HMD, Meta Quest 3 [55]. All participants received monetary compensation of $15 upon completing 70-minute sessions.

*5.1.3 Data Collection and Analysis.* We collected multiple data streams: audio recordings and transcripts of ideas, screen captures of visual experience, spatial movement tracking within VR, and time spent in different categories or global views. Ideas were categorized by two trained evaluators and assessed for creativity using fluency, flexibility, persistence, and originality metrics [28]. Inter-rater reliability was high for categorization (Cohen's κ = 0.774) and moderate for originality (Cohen's κ = 0.543). Both criteria were settled after discussion. Based on evaluated categories, we measured the context switching during ideation: a two-minute session where ideas 1–2 belong to Category A, ideas 3–5 to Category B, and idea 6 returns to Category A, results in two switches (A → B and B → A), averaging to one switch per minute. Based on embodied cognition research suggesting that spatial scale shapes cognitive processing [15, 17], we hypothesized distinct patterns for each condition:

- **H1:** Mini-scale overview will promote broader exploration (higher fluency and flexibility).
- **H2:** Body-scale exploration will promote deeper exploration (higher persistence).
- **H3:** Mini-scale overview will show more frequent context switching due to the strategic viewpoint.

For the BODY condition, we analyzed *spatial-semantic correspondence* by measuring the percentage of ideas that semantically matched the category of the location (island) where they were generated. For interview analysis, one author conducted a thematic analysis [8]. The author read the entire interview transcripts and open-coded them. Four authors participated in the multiple rounds of discussions, the authors reviewed the codes, identified recurring patterns, outlined themes, and revealed the impacts of different interfaces.

### 5.2 Results

*5.2.1 Impacts on Ideation Patterns.*

*Ideation Output.* Regarding H1 and H2, no statistically significant differences were observed between the MINI and BODY conditions for total sessions (see Table 2). Yet, temporal analysis revealed distinct patterns. In the first half of ideation sessions, MINI supported a rapid and broad exploration of ideas. Participants proposed 33% more ideas ($W$ = 3.5, $p$ < 0.01) and 31% more categories ($W = 4.0$, $p < 0.05$). During the second half, BODY showed numerically greater persistence of 1.66 compared to 1.46 of MINI, yet with no statistical significance. However, it bolstered the ideation consistently, showing higher fluency and flexibility compared to MINI in the second half.

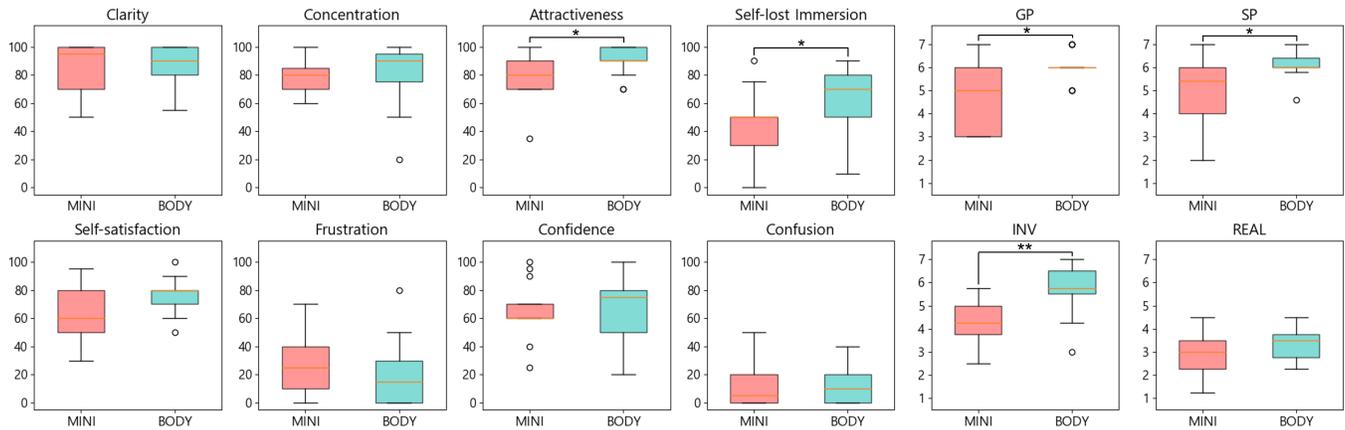

Figure 6: Experience evaluation results of Study 1 comparing MINI and BODY conditions. Left panels show general experience ratings on 0-to-100 scales: Clarity, Concentration, Attractiveness, Self-lost Immersion, Self-satisfaction, Frustration, Confidence, and Confusion. Right panels show spatial presence measurements on 7-point Likert scales: General Presence (GP), Spatial Presence (SP), Involvement (INV), and Experienced Realism (REAL). * indicates significance of $p < 0.05$, ** indicates significance of $p < 0.01$.

Table 2: Ideation outcome evaluation statistics of Study 1. The parentheses denote standard deviations. Cells highlighted in color indicate significant differences (orange: $p < 0.01$, yellow: $p < 0.05$) compared to the cells in the same columns, highlighted in gray.

| Condition | Fluency (# of ideas) | Flexibility (# of categories) | Persistence (with-in fluency) | Originality |
|---|---|---|---|---|
| **Total** | | | | |
| MINI | 10.08 (4.94) | 4.38 (1.39) | 2.23 (0.75) | 1.62 (1.61) |
| BODY | 9.38 (4.37) | 4.54 (0.88) | 2.02 (0.81) | 2.00 (1.41) |
| **First Half** | | | | |
| MINI | 5.62 (2.63) | 3.54 (1.51) | 1.56 (0.44) | 0.77 (0.93) |
| BODY | 4.23 (1.79) | 2.69 (1.11) | 1.63 (0.72) | 0.77 (1.09) |
| **Second Half** | | | | |
| MINI | 4.46 (2.76) | 2.92 (1.44) | 1.46 (0.44) | 0.85 (0.90) |
| BODY | 5.15 (2.97) | 3.15 (1.21) | 1.66 (0.76) | 1.23 (1.01) |

*Ideation Time and Context Switch Frequency.* The total session time showed that in MINI, participants had a mean ideation time of 311.9 seconds (SD=82.7), while in BODY, they recorded a longer duration with a mean of 339.4 seconds (SD=64.4). For H3, MINI had a higher mean frequency of 1.43 switches per minute (SD=0.83) compared to 1.07 switches per minute (SD=0.48) in BODY. The difference in switch frequency was statistically significant ($W = 21.0$, $p < 0.05$).

5.2.2 *Experience Evaluation.* BODY was preferred by 7 out of 13 participants for overall experience. For productivity, 8 participants preferred MINI, while 12 participants preferred BODY for immersiveness and interest. Figure 6 presents comprehensive experience evaluation results across 12 dimensions, revealing significant advantages for the BODY condition in both general experience and spatial presence measures.

*General Experience Ratings.* On the 0-to-100 experience scales, BODY demonstrated higher ratings in key experiential dimensions. Most notably, BODY yielded significantly higher attractiveness ratings ($W = 1.5$, $p < 0.05$) and self-lost immersion ($W = 17.5$, $p < 0.05$) compared to MINI.

*Spatial Presence Measurements.* The spatial presence evaluation revealed particularly strong effects favoring BODY across multiple dimensions. Participants reported significantly stronger *general presence* (overall sense of "being there"; $W = 2.5$, $p < 0.05$), enhanced *spatial presence* (feeling of being in the virtual environment's space; $W = 8.0$, $p < 0.05$), and notably higher *involvement* (engagement and focus with the virtual environment; $W = 1.5$, $p < 0.01$). *Experienced realism* (believability of the virtual environment) showed no significant difference between conditions. The fantasy-like visual aesthetic and the thought-driven nature of the environment led many participants to characterize both experiences as distinct from reality.

5.2.3 *Scale-Dependent Ideation Patterns.* Qualitative analysis revealed that the two interfaces fostered distinct ideation patterns and reward perceptions. In MINI, participants adopted a detached, "god-like" perspective, focusing on the proposal of new categories and fair distribution of ideas across the entire space. For example, S1P2 rapidly established multiple conceptual territories in the first half of his session (refer to Figure 9.a in Section A.2) by creating [Nutrition Management], [Exercise Promotion], and [Health Monitoring] categories. After scanning the emerging landscape, he identified gaps and filled underrepresented areas, adding [Health Education] through "teaching healthy lifestyle classes" and [Rewards & Incentives] by proposing "incentive programs for breaking unhealthy habits." This behavior reflected the primary motivation to create new islands, filling the empty conceptual landscape through deliberate gap identification and strategic placement.

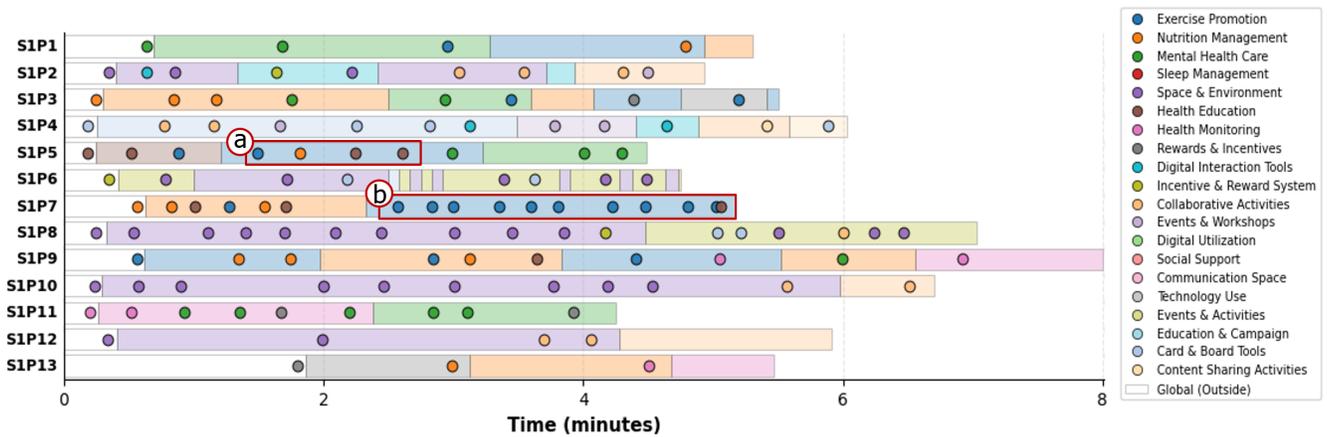

**Figure 7: The ideation and spatial dwelling patterns in Study 1 body-scale condition. Ideas are represented as colored circles with LLM-assigned categories, and user-dwelled areas are shown as lighter-colored regions. White indicates non-initiated (empty view) periods. The legend reflects two counterbalanced ideation tasks. Note that similar category names recur by different ideation sessions.**

In contrast, BODY promoted a focused, immersive experience. The physical act of walking to transition between categories made switching a deliberate, high-cost action. This encouraged users to deepen their thoughts and exhaust ideas within their current location before moving. For instance, S1P7 spent over two minutes dwelling within the [Exercise Promotion] island (see Figure 7.b), systematically exploring multiple facets of student fitness support. She began with infrastructure-focused ideas: "affordable sports programs at university facilities" and "low-cost access to futsal courts, yoga rooms, and gyms." Remaining within the same territory, she shifted to programmatic approaches: "sports competitions with other students," "mandatory physical education for freshmen," and "experience opportunities for unique sports like fencing and archery." Finally, she explored community and support mechanisms: "running crews and climbing clubs," "funding for group exercise applications," and "dedicated jogging courses and exercise spaces." In the post-session interview, S1P7 mentioned "Being on the same island and thinking about ideas in the same category allowed me to do more detailed ideation about that specific category." Participants also felt like "active protagonists in an adventure," and the act of walking itself influenced cognition; S1P4 noted that "circling around [the island] led to a deeper ideation process." The primary reward shifted from creating islands to populating the immediate environment with *large* trees, making the experience more game-like and relaxing.

*5.2.4 Spatial-Semantic Correspondence Analysis.* Across all BODY condition sessions, participants generated 122 total ideas, with 109 ideas (89.3%) produced within island environments (refer to Figure 7). Among the 109 categorized ideas generated within islands, 53 ideas semantically matched their spatial location, yielding a 48.6% spatial-semantic correspondence rate. The remaining 51.4% of mismatches included different interpretations with user-intents, ideas generated while moving toward another domain, and cross-category ideation. For example, S1P5 demonstrated cross-category thinking (see Figure 7.a), while dwelling within [Exercise Promotion] after proposing "increasing interest in sports clubs." She proposed "providing balanced meal plans," which subsequently created the new category [Nutrition Management]. As she moved closer to examine the newly formed orb, the [Health Education] orb entered her field of view, prompting her to generate "promotion through celebrities and influencers that students are interested in" and "adding regular health classes to the school curriculum"—both education-focused ideas while visually engaging with multiple category representations simultaneously.

*5.2.5 Dynamic Evolution Quality.* The dynamic evolution quality was evaluated in three dimensions: class naming, classification quality, and in-class consistency on a 0-to-100 scale. The MINI condition scored 86.3, 84.2, and 84.8, while the BODY scored 83.5, 89.6, and 88.9, respectively.

# 6 STUDY 2: TRANSITION WITHIN AND BETWEEN MODES

This study examines how users build creative workflows in an integrated dual-scale system. We focus on how participants leverage *dive-in* and *dive-out* transitions to move fluidly between overview and immersion, revealing natural exploration strategies.

## 6.1 Methodology

*6.1.1 Study Condition.* We used a single condition to determine the effects of dual-scale integration:

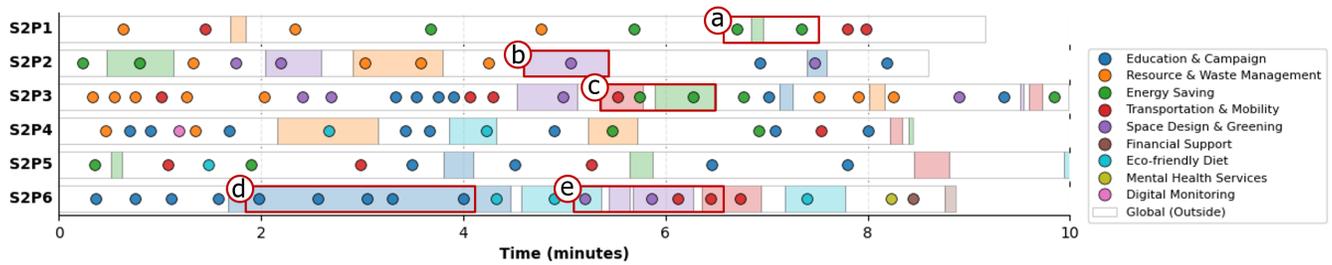

**Figure 8: The integrated navigation patterns in Study 2. Ideas are represented as colored circles with LLM-assigned categories, and user-dwelled areas are shown as lighter-colored regions. White indicates mini-scale overview periods.**

- **Inter-Scale Transition:** Provides seamless transitions between mini-scale overview and body-scale exploration (see *dive-in* and *dive-out* for details in Section 3.1.3).

*6.1.2 Participants and Procedure.* We recruited six participants from Study 1 (5 females; age mean=24.33, SD=3.78; see Section A.1 for detailed demographics). All 6 participants regarded them as proficient in ideation, marking them 4+ points on a 7-point Likert scale. At the start of the experiment, participants received task instructions and were introduced to using the artifact. They were requested to use the body-scale mode at least once during the session. Participants engaged in 8-10 minute ideation sessions on campus sustainability initiatives. They were notified at the 8-minute mark that they could continue or finish. When participants generated nine or more ideas within a single category, they were guided to explore other categories. This guidance was provided to one participant (S2P6). After the ideation, participants rated the system usability [9]. We conducted a semi-structured interview. The interview centered on the mode switch experience and how they used each mode. The experiment took place in a space larger than $33m^2$ where participants could move around safely. The experiment employed a VR HMD, Meta Quest 3 [55]. All participants received monetary compensation of $15 upon completing 70-minute sessions.

*6.1.3 Data Collection and Analysis.* We collected multiple data streams: audio recordings and transcripts of ideas, screen captures of visual experience, spatial movement tracking within VR, and categorical dwellings throughout the session. For ideas generated during body-scale exploration, we measured *spatial-semantic correspondence* (see definition in Section 5.1.3). The interview analysis followed the same protocol in Section 5.1.3, utilizing the thematic analysis [8].

## 6.2 Results

*6.2.1 Usage Patterns.* With the freedom to switch scales, participants developed distinct workflows, spending three quarters of their time (73.4%) in the mini-scale overview. The remaining 26.6% was dedicated to body-scale exploration, which accounted for 28% of generated ideas.

*6.2.2 Inter-Scale Transition Experience.* The ability to seamlessly switch modes gave participants a strong sense of control over their cognitive process. This enhanced control, combined with fast transitions, maintained or even increased immersion compared to the previous body-scale condition. S2P3 exemplified this fluid control (see Figure 8.c) by diving into [Transportation & Mobility] and proposing "reducing campus elevator usage," confirming the generated summary ("elevator usage reduction"), then immediately exclaiming as a new energy-related idea emerged: "creating a team to manage excessive heating and air conditioning." Recognizing this idea would not fit the current transportation category, she quickly dove out to overview before the idea materialized as a tree. Upon seeing the newly created [Energy Conservation] island, she expressed delight and dove in while repeating "energy conservation," then proposed "improving energy efficiency by identifying and upgrading poorly performing glass windows." This sequence demonstrated how rapid inter-scale transitions enabled users to match their spatial environment to their evolving thought processes in real-time. Most participants (5/6) saw the removal of the imposed walk for domain transition as an improvement, as it allowed for quick context-checking without disrupting their flow of thought. Still, two participants highlighted that the reduced physical walking diminished the sense of spatial engagement and relaxation they experienced previously. S2P2 remarked that the effortful walking in the previous condition "indeed reinforced the sensation of moving to a different category," fostering a more intense focus. This suggests that while seamless transitions benefit cognitive flow and control, the optimal level of physical embodiment may vary based on individual preferences.

*6.2.3 Analysis of Body-Scale Utilization.* Qualitative analysis revealed that participants leveraged the body-scale mode for three distinct purposes, demonstrating its versatile role in the creative process.

*Responsive Thinking Places.* For deep, contemplative ideation, participants used islands as responsive thinking environments. This pattern was characterized by longer stays (>40s) and sustained, often circular, movement within an island. For example, S2P6 dove into the [Education & Campaign] island and spent extended time developing interconnected ideas that progressively broadened the scope of educational engagement (see Figure 8.d). She began with "sustainability policy competitions," then expanded to social dimensions with "peer mentors for campus community building." Walking within the island, she further extended to international perspectives: "exchange student and study abroad programs" and "inter-institutional exchange programs with other countries and schools." Finally, she addressed internal relationships with "enhanced faculty-student interactions." This sequence demonstrated how sustained dwelling within a single conceptual territory enabled users to systematically explore multiple facets—from policy to social to international to interpersonal dimensions—creating a comprehensive ideation framework that might not emerge through rapid category switching.

*Information Filtering for Concentration.* Participants strategically used the body-scale mode to filter out the cognitive load of the overview. When feeling overwhelmed by the number of categories visible in the mini-scale, they would transition into a single island to focus. S2P2 demonstrated this pattern by diving into the [Space Design & Greening] island when the overview became cluttered with multiple categories (see Figure 8.b). She spent 30 seconds scanning the surroundings, then stood still in contemplation before proposing "college-specific garden design competition." After waiting for the system to generate the tree and label, she examined the summary of her newly created idea and reviewed other existing tree labels within the island before diving out. These visits were shorter (20-40s) but served as efficient, targeted sessions to mitigate cognitive pressure and concentrate on a single conceptual domain.

*Zooming to Distant or Occluded Content.* The body-scale mode also served a practical purpose for spatial inspection. Participants used it for brief (10-20s) "zoom-in" actions to get a clearer view of islands that were distant or whose content (trees) had become too dense to assess from the overview. S2P1 exemplified this (see Figure 8.a) by viewing the [Energy Conservation] category label from a distance and proposing "applying ink-saving printing methods for campus shared printers." When the idea generated as a tree on the distant island—the farthest of three visible islands that would require walking around islands to reach—he chose the efficient dive-in approach instead. He quickly dove into the island, verified the summarized content on the tree label, then dove out and remained in mini-scale to propose "building thermal design" while maintaining his strategic overview position. This allowed for quick, targeted content verification without losing the overall strategic context.

*6.2.4 Spatial-Semantic Correspondence Analysis.* Participants generated 93 total ideas across the dual-scale integration sessions, with 26 ideas (28.0%) produced during body-scale exploration within islands (refer to Figure 8). Among the 26 ideas generated within islands, 19 semantically matched their spatial location, yielding a 73.1% spatial-semantic correspondence rate. The remaining 26.9% of mismatches included different interpretations with user-intents and instantiation of a new island to move on. For instance, while dwelling within [Eco-friendly Diet], S2P6 proposed a study environment idea that created [Space Design & Greening] (see Figure 8.e). Rather than continuing in overview mode, she immediately dove out and into the newly formed island to maintain immersive ideation. This pattern repeated when she proposed a transportation-related idea that generated [Transportation & Mobility], again choosing to dive into the new category rather than remain in overview. This sequence illustrated how users' preference for embodied exploration could produce spatial-semantic misalignment.

*6.2.5 System Usability.* The system achieved high usability (SUS: 89.2), confirming that the integrated design supported an effective creative process.

# 7 DISCUSSION
## 7.1 Validating Explorable Ideas Framework

Our empirical findings provide comprehensive validation for the Explorable Ideas framework by demonstrating how spatial-semantic coordination enables effective spatial information exploration across different interaction modalities and user preferences.

*7.1.1 Spatial-Semantic Coordination through Dual-Scale Interfaces.* Our dual-scale investigation directly addresses RQ1 by revealing how different spatial scales within the Idea Islands system support spatial-semantic coordination through distinct cognitive engagement patterns. This validates the Explorable Ideas framework's core principle that embodied spatial exploration can meaningfully correspond to conceptual exploration [42, 88]. Overview perspectives foster reflective assessment—a strategic mode where users survey the entire conceptual landscape to identify patterns and plan exploration routes. This elevated perspective aligns with construal level theory's prediction that psychological distance promotes abstract, strategic thinking [86], with mini-scale showing 33% higher fluency and 31% higher flexibility in early ideation phases. The spatial positioning activates strategic thinking modes, providing comprehensive landscape awareness that supports planning and pattern recognition across entire ideation contexts [77]. Immersive perspectives enable persistent engagement—where embodied presence within conceptual territories transforms ideation from rapid category-switching to sustained dwelling within focused domains. Body-scale exploration's longer session times (339.4 vs. 311.9 seconds) and lower context switching frequency (1.07 vs. 1.43 switches per minute) reveal how physical movement constraints guides creative exploration. The embodied constraint of deliberate movement between categories transforms conceptual switching from a cognitive decision into a physical commitment, naturally encouraging thorough exploration of current domains and enabling multi-layered conceptual development within single territories [46]. The critical insight for the Explorable Ideas framework is that different spatial scales provide cognitive-spatial coordination rather than mere interface variety. Scale transitions become meaningful when they align with different cognitive needs, creating natural rhythms for creative work through complementary cognitive modes. This establishes that the framework's requirement for multi-perspective exploration can successfully support varied information engagement patterns.

*7.1.2 Representational Considerations in Spatial-Semantic Environments.* Two studies reveal fundamental representational requirements that enable effective spatial creativity support within the Explorable Ideas framework. The success of our islands-and-trees approach validates that natural containment relationships can effectively support semantic organization [42]. Aesthetic motivational affordances revealed that spatial artifacts must provide intrinsic satisfaction for continued exploration [99]. Participants' desire to "fill empty space with props generated through ideation" demonstrates how explorable ideas can transform abstract conceptual development into concrete spatial achievement, creating sustainable motivation for creative work. This extends the framework's representational requirements by demonstrating that spatial territories must not only organize semantic content but also provide motivational incentives for continued exploration.

*7.1.3 Embodiment Balance in Spatial Navigation.* Our investigation of navigation approaches reveals fundamental tensions between physical engagement and cognitive efficiency in embodied creativity systems. This analysis directly informs the framework's design consideration of how to support transitions within and between spaces while maintaining spatial-semantic coordination. Two distinct forms of immersion regarding navigation emerged from our studies. Walk-to-teleport created *embodied immersion*—presence derived from physical effort and environmental dwelling where participants valued the "relaxation" of walking through conceptual landscapes and felt like "active protagonists." However, dive-in and dive-out induced *transitional immersion*—a flow state characterized by smooth movement between cognitive modes and enhanced user control of environments. Study 2 results indicate a general preference for transitional over embodied immersion. Imposing physical movement to switch environment contexts can distort thought processes to fit bodily pace, hindering the natural rhythm of creative ideation. This suggests that physical movements should support rather than constrain users' cognitive preferences, with systems providing embodiment options while prioritizing cognitive agency.

## 7.2 Spatial-Semantic Correspondence as a Design Evaluation Framework

Through our analysis across two studies, we discovered that spatial-semantic correspondence (SSC) emerges as a valuable metric for evaluating spatial creativity support effectiveness within the Idea Islands probe design. Three lines of evidence establish SSC as a fundamental indicator for systems that map semantic contexts onto spatial domains. First, our contrasting SSC patterns reveal meaningful distinctions between system limitations and user agency: Study 1's lower correspondence (48.6%) reflected *system-imposed friction* through transitional misalignment, visual interference, and forced dwelling, while Study 2's higher correspondence (73.1%) enabled *user-driven exploration* where misalignments reflected intentional cross-category ideation rather than system constraints. Second, interview data demonstrates that SSC improvements overlap with users' perceived cognitive-spatial control, with Study 2 participants valuing their ability to "keep the thoughts flowing and connect ideas instantly" and "freedom of movement prompted new exploration." Third, behavioral analysis shows that users naturally allocate different cognitive functions to different scales when provided appropriate tools—selecting the mini-scale for cross-category ideations and the body-scale for focused dwelling—demonstrating that higher SSC reflects enhanced creative control rather than mere spatial compliance. This metric provides actionable feedback for system design by identifying whether spatial affordances successfully support conceptual exploration or create unnecessary friction that constrains creative agency. Note that we use AI-generated categorizations as the SSC evaluation baseline because users' spatial experiences are inherently shaped by real-time AI categorization, making alternative human categorizations misrepresentative of actual cognitive-spatial coordination.

SSC represents a fundamental evaluation framework for Explorable Ideas because it directly measures the core principle underlying spatial information exploration: the meaningful correspondence between physical navigation and conceptual discovery. This metric extends beyond our creativity support implementation to capture the essential quality of any system that spatializes abstract information, whether users' spatial activities reflect their cognitive exploration intentions. For knowledge exploration systems, SSC could evaluate the alignment between specific locations and the questions, manipulations, or activities users perform there, logging behavioral patterns to assess how well spatial positioning corresponds to conceptual engagement. This promises SSC as a generalizable indicator that can validate whether spatial metaphors successfully support users' natural exploration patterns.

## 7.3 Design Implications for Implementing Explorable Ideas

Our findings establish three foundational design implications for implementing the Explorable Ideas framework across different platforms and interaction modalities, extending beyond our VR implementation to inform broader spatial information exploration systems.

(1) **Calibrate exploration through prop aesthetics and scale.** Use visually appealing artifacts with carefully chosen sizes to shape users' sense of reward. Since users perceive prop instantiation from exploration as spatial territory filling, symbolic artifact sizes become critical variables that influence user preferences and guide next-step decisions within the reward system.
(2) **Balance embodied immersion with cognitive efficiency.** Leverage embodied navigation to foster engagement, but avoid excessive physical effort that disrupts semantic flow. Our findings reveal that forced embodied navigation can distort ideation patterns through cognitive burden, though users simultaneously experience engagement and satisfaction. Design lightweight navigation and mode-switching mechanisms that maintain spatial-semantic correspondence while preserving users' natural cognitive flow.
(3) **Utilize spatial-semantic correspondence (SSC) as an evaluation criterion.** Employ SSC as a key metric for assessing interaction design appropriateness, particularly for spatial context transitions and domain switching. The degree of alignment between spatial exploration and semantic exploration serves as a core indicator of users' creative agency and system effectiveness within the framework's spatial-semantic coordination paradigm.

## 7.4 Limitations and Future Works

While our studies establish the foundational design considerations for explorable ideas systems, several limitations point toward important future research directions. Our current implementation demonstrates proof-of-concept capabilities but reveals opportunities for advancing spatial information exploration through emerging technologies and more sophisticated interaction design.

*7.4.1 Dynamic Spatial Metaphor Generation.* Our current implementation relies on predefined spatial metaphors—islands, trees, and orbs—that remain static across different information contexts. This limitation becomes apparent when users engage with diverse domains that might benefit from more contextually-appropriate spatial representations.

Future systems could leverage 3D object generation technologies [43, 49, 64, 68] to create domain-specific metaphors: architectural concepts might generate building-like structures, while knowledge exploration could produce library-like environments. The key challenge involves maintaining spatial consistency while accommodating metaphor evolution, requiring systems that preserve users' spatial memory while introducing motivational variety for sustained engagement.

*7.4.2 Cross-Category Exploration Mechanisms.* Our hierarchical island-and-tree structure successfully supports focused exploration within individual categories but provides limited mechanisms for discovering connections across conceptual boundaries. This limitation restricts users' ability to explore inter-category relationships while maintaining the cognitive benefits of hierarchical organization.

Future implementations could address this through bridges that emerge between islands when users generate spanning ideas exploration, or portal-like gateways showing relative contents via peeking views or partial overlaps. The design challenge involves balancing focused exploration with serendipitous discovery without overwhelming users with excessive cross-category suggestions.

*7.4.3 Scale-Specific Interaction Design.* Our current navigation system treats both scales as functionally equivalent contexts, differing only in perspective. This limitation overlooks opportunities to provide scale-specific interaction capabilities that leverage the unique affordances of each mode for different creative functions.

Future implementations could support "god-like" interactions in mini-scale environments—picking up and rearranging entire islands or relocating individual trees between categories to support strategic reorganization. Body-scale environments could enable intimate interactions such as approaching trees for detailed displays, physical gestures to activate concept development prompts, or embodied editing capabilities where users physically manipulate spatial elements to refine categorizations. The key principle involves ensuring that interaction capabilities match the cognitive affordances of each scale while preserving users' creative momentum during transitions.

*7.4.4 Collaborative and Longitudinal Use.* Our single-user, short-session studies leave critical questions about how spatial representations should evolve across multiple users and extended time periods. This limitation constrains our understanding of how explorable ideas systems could support collaborative information exploration and long-term knowledge development.

Multi-user scenarios require *consensus-based spatial evolution* mechanisms where categorization disagreements create temporary *disputed territories* until collaborative resolution. Longitudinal usage presents challenges around *temporal coherence*, requiring graceful category migration without losing spatial memory as projects evolve over weeks or months. Advanced implementations could incorporate *predictive spatial adaptation* that anticipates users' evolving needs based on usage patterns. The fundamental challenge involves balancing spatial consistency for long-term mental model development with responsiveness to changing information needs across different domains and collaboration patterns.

These limitations collectively point toward opportunities for advancing spatial information exploration systems while preserving the core benefits of embodied navigation. Future research should investigate how emerging technologies can address these constraints while maintaining the intuitive spatial interaction that makes explorable ideas effective for diverse information exploration contexts.

## 8 CONCLUSION

This work establishes Explorable Ideas as a design paradigm for spatial creativity support, demonstrating that abstract ideation can be externalized into navigable environments that leverage embodied interaction. Through the Idea Islands probe and two controlled studies, we showed how overview and immersion perspectives provide complementary affordances and how integrated transitions allow users to fluidly combine them into effective workflows. Our findings highlight the potential of spatial environments to guide exploration, sustain engagement, and align conceptual development with embodied activity. The concept points toward new ways of designing systems that transform abstract information into lived, explorable experiences across creative, educational, and knowledge-intensive domains.

# A APPENDIX

## A.1 Participant Demographics

Table 3 presents detailed demographics and characteristics of all participants across Studies 1 and 2. Study 2 recruited six participants from the previous study, maintaining consistency in participant characteristics.

## A.2 Study 1 Mini-Scale Process Visualization

Figure 9 illustrates the ideation patterns observed in Study 1's mini-scale condition. The overview perspective enabled participants to adopt strategic, 'god-like' positioning behaviors, systematically identifying conceptual gaps and distributing ideas across the entire landscape to create balanced coverage of the ideation space.

Table 3: Participant demographics and characteristics for Studies 1 and 2. The self-reported ideation proficiency is rated on a 7-point Likert scale. VR usage experience is the number of prior VR experiences.

| Study 1 ID | Study 2 ID | Age | Gender | Education Level | Ideation Frequency | Ideation Proficiency | VR Usage Experience |
|---|---|---|---|---|---|---|---|
| S1P1 | S2P1 | 27 | M | Graduate | Less than monthly | 5 | 5+ |
| S1P2 | – | 26 | M | Undergraduate | 1-2 times daily | 6 | 5+ |
| S1P3 | S2P2 | 23 | F | Undergraduate | 1-2 times weekly | 6 | 5+ |
| S1P4 | S2P3 | 24 | F | Graduate | 1-2 times weekly | 5 | 3 |
| S1P5 | – | 19 | F | Undergraduate | 1-2 times weekly | 5 | 4 |
| S1P6 | S2P4 | 30 | F | Graduate | 1-2 times daily | 4 | 2 |
| S1P7 | – | 27 | F | Graduate | 1-2 times weekly | 4 | 2 |
| S1P8 | – | 24 | F | Undergraduate | 1-2 times weekly | 4 | 5+ |
| S1P9 | S2P5 | 19 | F | Undergraduate | Less than monthly | 6 | 3 |
| S1P10 | – | 25 | M | Undergraduate | Less than monthly | 1 | 3 |
| S1P11 | S2P6 | 23 | F | Undergraduate | 1-2 times weekly | 5 | 4 |
| S1P12 | – | 21 | M | Undergraduate | Less than monthly | 2 | 5+ |
| S1P13 | – | 22 | M | Undergraduate | Less than monthly | 2 | 2 |

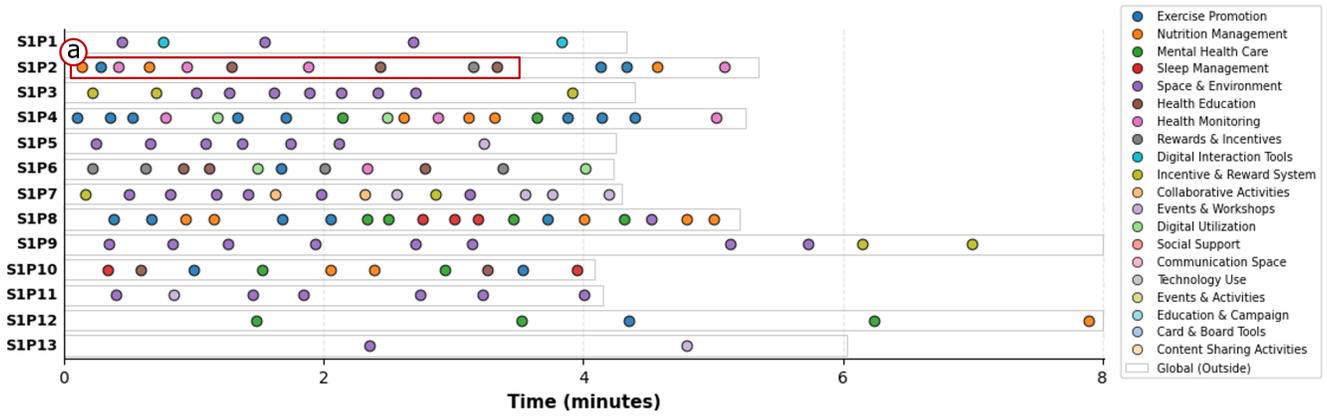

Figure 9: The ideation patterns in Study 1 mini-scale condition. Ideas appear as colored circles with LLM-assigned categories. The legend reflects two counterbalanced ideation tasks.

---

**Algorithm 1:** Dynamic Evolution of Visual Metaphors

**Input:** $S_{input}$: User's speech utterance
**Output:** Updated VR environment
**Data Structures:**
$T_{transcript}$: Speech-to-text transcription
$(C_{inferred}, Content_{summary})$: LLM-inferred category and idea summary
$Islands_{existing}$: Set of current category islands

```
// Step 1: Speech Transcription
```
$T_{transcript} \leftarrow \text{STT}(S_{input})$
```
// Step 2: Context-Aware LLM Inference
```
$(C_{inferred}, Content_{summary}) \leftarrow \text{LLM}(T_{transcript}, Context)$
```
// Step 3: Adaptive Island Management
```
**if** $C_{inferred} \notin Islands_{existing}$ **then**
  Create new island for $C_{inferred}$
  $Islands_{existing} \leftarrow Islands_{existing} \cup \{C_{inferred}\}$
**end**
```
// Step 4: Tree Generation and Placement
```
Add tree with $Content_{summary}$ to island $C_{inferred}$
```
// Step 5: Environment Rendering
```
Render with new/modified islands and trees

### A.3 LLM Prompt Design for Semantic Organization

The system employs few-shot prompting with LangChain's FewShotPromptTemplate to implement consistent semantic organization across different ideation contexts. The prompt templates use predefined category suggestions while allowing dynamic creation of new categories when existing ones are insufficient.

Figure 10 presents the prompt for Study 1's dual-topic ideation sessions supporting both *campus lounge/cafe communication enhancement* and *healthy living habit promotion*.

Figure 11 shows the prompt for Study 2's *sustainable campus* focus, maintaining the same structured approach with sustainability-specific categories.

```
_PREFIX = """
Your task is to categorize and summarize Korean transcripts about a campus innovation challenge.
Output must be a single-line string in the format: CATEGORY;SUMMARY
(CATEGORY and SUMMARY each in 1-3 words and 1-5 words, no extra explanation).

We have two main topics:
1) "Campus lounge/cafe communication" — how to foster everyday communication in an indoor lounge/cafe setting.
   Example categories to consider (in Korean):
   - Digital Interaction Tools
   - Card & Board Tools
   - Events & Workshops
   - Collaborative Activities
   - Content Sharing Activities
   - Space & Environment
   - Experience & Practice Activities
   - Incentive & Reward Systems
   (You may create a new category if none of the above apply.)
2) "Healthy living habits" — how to encourage students to adopt healthier daily routines.
   Example categories to consider (in Korean):
   - Nutrition Management
   - Exercise Promotion
   - Health Monitoring
   - Rewards & Incentives
   - Mental Health Care
   - Health Education
   - Digital Utilization
   (You may create a new category if none of the above apply.)

Rules:
1. Do not invent content not in the transcript.
2. If the transcript fits an existing category from the provided list, use it rather than creating a new one.
3. Only create a new category if the idea is clearly different from all existing categories.
4. Summaries must be accurate and preserve the important content (1-5 words; average 3~4 words).
5. Do not create meaningless categories such as "Other" or "New usage".
6. Output must be "CATEGORY;SUMMARY".

Double-check:
- If the transcript's idea is already covered by an existing category, do NOT create a new one.
- Avoid subdividing categories too finely unless truly necessary.
- Use only the content that actually appears in the transcript (no extra details).
"""
_EXAMPLE = [
    # 1) Lounge/cafe communication
    {
        "topic": "Ideas to promote communication in campus indoor lounge/cafe",
        "categories": ["Events & Workshops", "Card & Board Tools"],
        "transcript": "If we install a large touch screen in one corner of the lounge and hold weekly quiz events, I think conversations would naturally emerge.",
        "output": "Digital Interaction Tools;touch screen quiz events"
    },
    # 2) Lounge/cafe communication
    {
        "topic": "Ideas to promote communication in campus indoor lounge/cafe",
        "categories": ["Events & Workshops", "Collaborative Activities"],
        "transcript": "If we mix quiet spaces and open-type spaces half and half in the lounge, people could naturally chat or rest according to their preferences.",
        "output": "Space & Environment;quiet zones and open zones combined"
    },
    # 1) Healthy habits promotion
    {
        "topic": "Ways to encourage students' healthy living habits",
        "categories": ["Health Monitoring"],
        "transcript": "How about using an app to listen to meditation sounds together when stressed and give feedback to each other?",
        "output": "Mental Health Care;meditation app sharing for stress management"
    }
]
```

**Figure 10: The LLM prompt for Study 1 ideations.**

```
_PREFIX = """
Your task is to categorize and summarize Korean transcripts about a campus innovation challenge.
Output must be a single-line string in the format: CATEGORY;SUMMARY
(CATEGORY and SUMMARY each in 1-3 words and 1-5 words, no extra explanation).

We have one main topic:
"Sustainable campus" — how to improve sustainability within a university campus setting.

Example categories to consider (in Korean):
- Energy Saving
- Resource & Waste Management
- Transportation & Mobility
- Space Design & Greening
- Eco-Friendly Diet
- Education & Campaign
- Digital Monitoring
(You may create a new category if none of the above apply.)

Rules:
1. Do not invent content not in the transcript.
2. If the transcript fits an existing category from the provided list, use it rather than creating a new one.
3. Only create a new category if the idea is clearly different from all existing categories.
4. Summaries must be accurate and preserve the important content (1-5 words; average 3~4 words).
5. Do not create meaningless categories such as "Other" or "New usage".
6. Output must be "CATEGORY;SUMMARY".

Double-check:
- If the transcript's idea is already covered by an existing category, do NOT create a new one.
- Avoid subdividing categories too finely unless truly necessary.
- Use only the content that actually appears in the transcript (no extra details).
"""
_EXAMPLE = [
    # 1) Energy Saving example
    {
        "topic": "Creating sustainable campus environment",
        "categories": ["Resource & Waste Management", "Energy Saving"],
        "transcript": "The problem is that classroom lights and air conditioners are left on after work hours. How about expanding night patrol staff?",
        "output": "Energy Saving;night patrol waste prevention"
    },
    # 2) Digital Monitoring example
    {
        "topic": "Creating sustainable campus environment",
        "categories": ["Energy Saving"],
        "transcript": "If we visualize power and water usage by campus building in real-time and publish it on the web, I think people's conservation awareness would increase.",
        "output": "Digital Monitoring;real-time resource usage disclosure"
    }
]
```

**Figure 11: The LLM prompt for Study 2 ideations.**